\documentstyle[11pt]{article}
\begin{document}

\title{The effects of twisted magnetic field on coronal loops oscillations and
dissipation}
\author{K. Karami$^{1,2}$,\thanks{E-mail: KKarami@uok.ac.ir}\\
M. Barin${^1}$,\thanks{E-mail: Barin$_{_{-}}$m$_{_{-}}$2046@yahoo.com  }\\
$^{1}$\small{Department of Physics, University of Kurdistan,
Pasdaran St., Sanandaj, Iran}\\$^{2}$\small{Research Institute for
Astronomy $\&$ Astrophysics of Maragha (RIAAM), Maragha, Iran}}

\maketitle

\begin{abstract}
The standing MHD modes in a zero-$\beta$ cylindrical magnetic flux
tube modelled as a straight core surrounded by a magnetically
twisted annulus, both embedded in a straight ambient external field
is considered. The dispersion relation for the fast MHD waves is
derived and solved numerically to obtain the frequencies of both the
kink ($m=1$), and fluting ($m=2,3$) waves. Damping rates due to both
viscous and resistive dissipations in presence of the twisted
magnetic field is derived and solved numerically for both the kink
and fluting waves.
\end{abstract}


\section{Introduction}
\label{intro} Solar corona is a highly structure of magnetic flux
tubes namely coronal loops. Transverse oscillations of coronal loops
were first identified by Aschwanden et al. (1999) and Nakariakov et
al. (1999) using the observations of TRACE. Edwin \& Roberts (1983)
elaborated on the dispersion relation for a magnetic cylinder
embedded in a magnetic environment typical of that of the solar
photosphere and corona. They found that the existence of
inhomogeneities in the form of structuring of the magnetic field
enables loops to act as wave guides for a variety of different
modes. Karami, Nasiri $\&$ Sobouti (2002) used the model of Edwin \&
Roberts (1983) but without limiting it to slender tubes. They solved
numerically the dispersion relation for each mode in its full
generality. They obtained that in presence of weak viscous and ohmic
dissipations, the damping rate is inversely proportional to the
Reynolds and Lundquist numbers, R and S, respectively.

An additional features of the flux tube is that
 of twist. Bennett, Roberts $\&$ Narain (1999) examined the influence of magnetic twist on the modes of
 oscillations of a magnetic flux tube. They found that twist
 introduces an infinite band of body modes. Klimchuk, Antiochos $\&$ Norton (2000) introduced twist to resolve the internal
 structure on an individual loop embedded within a much larger dipole
 configuration. Mikhalyaev $\&$ Solov'ev (2005) investigated the MHD waves
 in a double magnetic flux tube embedded in a uniform external
 magnetic field. The tube consists of a dense hot cylindrical
 cord surrounded by a co-axial shell. They found two slow and two
 fast magnetosonic modes can exist in the thin double tube.

Verwichte et al. (2004), using the observations of TRACE, detected
the multimode oscillations for the first time. They found that two
loops are oscillating in both the fundamental and the first-overtone
standing fast kink modes. According to the theory of MHD waves, for
uniform loops the period ratio $P_1/2P_2$ of the fundamental mode
and its first overtone is exactly 1. But the ratios found by
Verwichte et al. (2004) are 0.91$\pm$0.04 and 0.79$\pm$0.03 and thus
clearly differ from 1. This may be caused by different factors such
as the effects of curvature, see e.g. Van Doorsselaere et al.
(2004), leakage, see De Pontieu et al. (2001), density
stratification in the loops, see e.g. Andries et al. (2005), Erdelyi
\& Verth (2007), Karami \& Asvar (2007) and magnetic twist, see
Erd\'{e}lyi \& Fedun (2006) and Erd\'{e}lyi \& Carter (2006).

 Erd\'{e}lyi \& Fedun (2006), studied the wave propagation in twisted cylindrical
 magnetic flux tube embedded in an incompressible but
 also magnetically twisted plasma. They found that magnetic twist
 will increase, in general, the periods of waves approximately by
 a few percent when compare to their untwisted
 counterparts. Erd\'{e}lyi \& Carter (2006) used the model of Mikhalyaev \&
 Solov'ev (2005) but for a fully magnetically twisted
 configuration consisting of a core, annulus and external region.
 They investigated their analysis by considering magnetic twist
 just in the annulus, the internal and external regions having
 straight magnetic field. Two modes of oscillations occurred in
 this configurations; surface and hybrid modes. They found that
 when the magnetic twist is increase the hybrid modes cover a wide
 range of phase speeds, centered around the annulus, longitudinal
 Alfv\'{e}n speed for the sausage modes.

 Carter \& Erd\'{e}lyi (2007) investigated the oscillations of a magnetic flux tube
 configuration consisting of a core, annulus and external region
 each with straight distinct magnetic field in an incompressible medium. They found that there
 are two surface modes arising for both the sausage and kink modes
 for the annulus-core model where the monolithic tube has solely one surface mode for the incompressible
 case. Also they showed that the existence and width of an annulus
 layer has an effect on the phase speeds and periods. Carter \& Erd\'{e}lyi (2008) used the model introduced by Erd\'{e}lyi \&
 Carter (2006) to include the kink modes. They found for the set of
 kink body modes, the twist increase the phase speeds of the modes. Also they showed that there are two surface modes
 for the twisted shell configuration, one due to each surface, where
 one mode is trapped by the inner tube, the other by the annulus
 itself.

In the present work, our aim is to investigate the effects of the
twisted magnetic field on oscillations and damping of standing MHD
waves in the cold coronal loops observed by Verwichte et al. (2004)
deduced from the TRACE data. This paper is organized as follows. In
Section 2 we use the model introduced by Erd\'{e}lyi \&
 Carter (2006) to derive the equations of motion, introduce the
relevant boundary conditions and obtain the dispersion relation. In
Section 3 we discuss resistive and viscous dissipations to calculate
contributions of the different modes to heating of the coronal
loops. In Section 4 we give numerical results. Section 5 is devoted
to conclusions.

\section{Equations of Motion}
The linearized MHD equations for a zero-$\beta$ incompressible
plasma are
\begin{eqnarray} \frac{\partial\delta\mathbf{v}}{\partial
t}=\frac{1}{4\pi\rho}\{(\nabla\times\delta\mathbf{B})\times\mathbf{B}
+(\nabla\times\mathbf{B})\times\delta\mathbf{B}\}
+\frac{\eta}{\rho}\nabla^2\delta\mathbf{v},\label{mhd1}
\end{eqnarray}
\begin{eqnarray}
\frac{\partial\delta\mathbf{B}}{\partial
t}=\nabla\times(\delta\mathbf{v}\times\mathbf{B})+
\frac{c^2}{4\pi\sigma}\nabla^2\delta\mathbf{B},\label{mhd2}
\end{eqnarray}
\begin{eqnarray}
P_{\rm T}=\frac{\mathbf{B}\cdot\delta\mathbf{B}}{4\pi},\label{mhd3}
\end{eqnarray}
\begin{eqnarray}
\nabla\cdot\delta\mathbf{v}=0,\label{mhd4}
\end{eqnarray}
where $\delta\bf{v}$, $\delta\bf{B}$ and $P_{\rm T}$ are the
Eulerian perturbations in the velocity, magnetic field and magnetic
pressure, respectively; $\rho$, $\sigma$, $\eta$ and $c$ are the
mass density, the electrical conductivity, the viscosity and the
speed of light, respectively. Note that Eq. (\ref{mhd4}) satisfies
the incompressibility condition.

The simplifying assumptions are:
\begin{itemize}
\item $\rho$ is constant over the loop;
\item for a zero-$\beta$ loop, gas pressure is negligible;
\item from Erd\'{e}lyi $\&$ Carter (2006), background magnetic field is assumed to be
$$
{\mathbf{B}}=\left\{\begin{array}{lll}
{\mathbf{B}}_i=(0,0,B_i),~~~~~~~r<a,\\
{\mathbf{B}}_0=(0,A_0r,B_0),~~~a<r<R,\\
{\mathbf{B}}_e=(0,0,B_e),~~~~~~~r>R,\\
\end{array}\right.
$$
where $A_0$, $B_i$, $B_0$, $B_e$ are constant and $a$, $R$ are radii
of the core and tube, respectively;
      \item tube geometry is a circular with cylindrical
coordinates, ($r,\phi,z$);
     \item there is no initial steady flow over the tube;
\item viscous and resistive coefficients, $\eta$ and
$\sigma$ respectively, are constants;
\item t-, $\phi$- and z-  dependence for any of the components
$\delta{\bf{v}}$ and $\delta{\bf{B}}$ is $\exp{\{i(m\phi+k_z
z-\omega t)\}}$. Where $k_{z}=l\pi/L$, $L$ is length of the tube,
and $l=(1,2,\cdot\cdot\cdot)$, $m=(0,1,2,\cdot\cdot\cdot)$ are the
longitudinal and azimuthal mode numbers, respectively.
   \end{itemize}

We will further assume that the dissipative terms in Eqs.
(\ref{mhd1}) and (\ref{mhd2}) are much smaller. We will first solve
the problem without these terms and re-introduce them later as small
corrections in calculating contributions of the different modes to
heating of the corona. Taking time derivative of Eq. (\ref{mhd2})
and substituting for $\partial \delta {\mathbf{v}}/\partial t$ from
Eq. (\ref{mhd1}), the resulting equation yields to Bessel's equation
\begin{eqnarray}
\frac{d^2{P_{\rm T}}}{dr^2}+\frac{1}{r}\frac{d{P_{\rm
T}}}{dr}-\Big(\frac{m^2}{r^2}+m_{0}^2\Big){P_{\rm
T}}=0,\label{Bessel}
\end{eqnarray}
where
\begin{eqnarray}
m_{0}^2=k_{z}^2\Big[1-\frac{{A_{0}^2}{\omega_{A_0}^2}}{\pi\rho_{0}({\omega^2}-{\omega_{A_0}^2})^2}\Big],
\end{eqnarray}
and
\begin{eqnarray}
{\omega_{A_0}}=\frac{1}{\sqrt{4\pi\rho_0}}(mA_0+k_{z}B_0).
\end{eqnarray}
Equation (\ref{Bessel}) is same as the result exactly derived by
Erd\'{e}lyi $\&$ Carter (2006). Note that subscripts 0 (which
correspond to annulus) are replaced by $i$, $e$ corresponding to the
internal and external regions, respectively. Since for the internal
and external regions $A_i$=$A_e$=0, hence $m_i^2=m_e^2=k_z^2>0$ and
${\omega_{A_i}}=\frac{k_{z}B_i}{\sqrt{4\pi\rho_i}}$,
${\omega_{A_e}}=\frac{k_{z}B_e}{\sqrt{4\pi\rho_e}}$.

Solutions of Eq. (\ref{Bessel}) are:
\begin{eqnarray}
P_{\rm T}=\alpha I_m{(k_zr)},
\end{eqnarray}
for the interior region ($r<a$),
\begin{eqnarray}
P_{\rm T}=\Big\{\begin{array}{ll}
\beta I_m(m_0r)+\gamma K_m(m_0r),~~m_0^2>0,\\
\beta J_m(n_0r)+\gamma Y_m(n_0 r),~~~~~n_0^2=-m_0^2>0,\\
 \end{array}
\end{eqnarray}
for the annulus region ($a<r<R$) and
\begin{eqnarray}
P_{\rm T}=\delta K_m{(k_zr)},
\end{eqnarray}
for the exterior region ($r>R$). Where ($J_m,Y_m$) and ($I_m,K_m$)
are the Bessel and modified Bessel functions of the first and second
kind, respectively. The coefficients $\alpha,\beta,\gamma$ and
$\delta$ are determined by the boundary conditions. From both
Karami, Nasiri $\&$ Sobouti (2002) and Erd\'{e}lyi \& Carter (2006),
the necessary boundary conditions are that: at the boundaries $r=a$
and $r=R$, both the Lagrangian magnetic pressure and $\delta v_r$
should be continuous. These conditions yield to the dispersion
relations for surface, $m_0^2>0$, and hybrid, $m_0^2<0$, modes which
are same as the results obtained by Erd\'{e}lyi \& Carter (2006) in
Eqs. (28a) to (28b), respectively. Note that numerical solution of
the dispersion relation yields to eigenfrequencies, which are
characterized by a trio of wave numbers ($n,m,l$) that actually
count the number of nodes or antinodes along $r,~\phi$, and $z$
directions, respectively.

\section{Dissipative Processes}
Since the discovery of the hot solar corona about 66 years ago,
different theories of coronal heating have been put forward and
debated. For instance, Nakariakov et al. (1999) reported the
detection of spatial oscillations in five coronal loops with periods
ranging from 258 to 320 s. The decay time was $(14.5\pm2.7)$ minutes
for an oscillation of $(3.9\pm0.13)$ millihertz. Also Wang \&
Solanki (2004) described a loop oscillation observed on 17 April
2002 by TRACE in 195${\AA}$. They interpreted the observed loop
motion as a vertical oscillation, with a period of 3.9 minutes and a
decay time of 11.9 minutes. All these observations indicate strong
dissipation of the wave energy that may be the cause of coronal
heating.

Following Karami \& Asvar (2007), the finite conductivity and
viscosity of plasma causes an exponential time decay of
disturbances. Hence for weak dissipations one may assume
\begin{eqnarray}
\delta{{\mathbf{B}}}^{\rm
dissipative}&=&\delta{\mathbf{B}}({\mathbf{r}})e^{-(i\omega
+\alpha)t},
\nonumber\\
\delta{{\mathbf{v}}}^{\rm
dissipative}&=&\delta{{\mathbf{v}}}({\mathbf{r}})e^{-(i\omega
+\alpha)t}, \label{diss1}
\end{eqnarray}
 where $\omega$, $\delta{\bf{B}}$, and
$\delta{\bf{v}}$ on the right hand side are the solutions of Eqs.
(\ref{mhd1}), (\ref{mhd2}) in the absence of dissipations.
Substituting Eq. (\ref{diss1}) in Eqs. (\ref{mhd1}) and
(\ref{mhd2}), canceling out the non dissipative terms, and keeping
only the first order terms in $\alpha$, $c^2/4\pi\sigma$ and $\eta$
gives

\begin{eqnarray}
2i\alpha\omega\delta{\mathbf{B}}&=&\frac{c^2}{4\pi\sigma}\Big\{\nabla^{2}[(\mathbf{B}\cdot\nabla)\delta\mathbf{v}]-\nabla^{2}[(\delta\mathbf{v}\cdot\nabla)\mathbf{B}]\Big\}
\nonumber\\&+&\frac{\eta}{\rho}\Big\{(\mathbf{B}\cdot\nabla)\nabla^{2}\delta\mathbf{v}-(\nabla^{2}\delta\mathbf{v}\cdot\nabla)\mathbf{B}\Big\},\label{alphaEq}
\end{eqnarray}
where
\begin{eqnarray}
\delta\mathbf{v}&=&-\frac{i}{4\pi\rho}\frac{\omega}{(\omega^2-\omega_{A}^2)}\Big\{\mathbf{\nabla}(\mathbf{B}\cdot\delta\mathbf{B})-(\delta\mathbf{B}\cdot\mathbf{\nabla})\mathbf{B}\Big\}.
\end{eqnarray}

Rewriting Eq. (\ref{alphaEq}) for either the transverse or the
z-component and substituting for all quantities in terms of
$\delta{B}_{z}$ gives
\begin{eqnarray}
\alpha&=&\Big(\frac{k_z^2-m_0^2}{2}\Big)\Big[\frac{c^2}{4\pi\sigma}+\frac{\eta}{\rho_0}\Big],\nonumber\\
&=&\frac{v_{A_i}{\rm
R}}{\rho_0}\Big(\frac{\omega_{A_0}k_{z}A_{0}}{2\pi(\omega^2-\omega_{A_0}^2)}\Big)^2\Big[\frac{1}{S}+\frac{1}{\mathcal{R}}\Big],\label{alpha}
\end{eqnarray}
where $v_{A_i}=\frac{B_i}{\sqrt{4\pi\rho_i}}$, the Lundquist number
$S=\Big(\frac{4\pi\sigma {\rm R}^2}{c^2}\Big)/\Big(\frac{2\pi {\rm
R} }{v_{A_i}}\Big)$, is the ratio of the resistive time scale to the
Alfv\'{e}n crossing time and the Reynolds number
${\mathcal{R}}=\Big(\frac{{\rm
R}^2\rho_0}{\eta}\Big)/\Big(\frac{2\pi {\rm R}}{v_{A_i}}\Big)$ is
the ratio of the viscous time scale to the Alfv\'{e}n crossing time.
Equation (\ref{alpha}) shows that when the twist is absent, $A_0=0$,
the damping rate is vanished. Whereas for compressible plasma it is
not zero. See Karami, Nasiri $\&$ Sobouti (2002).
\section{Numerical Results}
As typical parameters for a coronal loop, we assume $L=109\times
10^3$ km, ${\rm R/L}=0.01$, $\rho_{e}/\rho_{i}=0.1$,
$\rho_{0}/\rho_{i}=0.5$, $\rho_{i}=2\times 10^{-14}$ gr cm$^{-3}$,
$B_{e}/B_{i}=1$, $B_{0}/B_{i}=1$, $B_{i}=100$ G. For such a loop one
finds $v_{A_{i}}=2000$ km s$^{-1}$, $\omega_{A_{\rm
i}}:=\frac{v_{A_{i}}}{{\rm R}}=1.835$ rad s$^{-1}$. We use $S=10^4$
and ${\mathcal{R}}=560$, given by Ofman et al. (1994).

\begin{figure}
\center \includegraphics{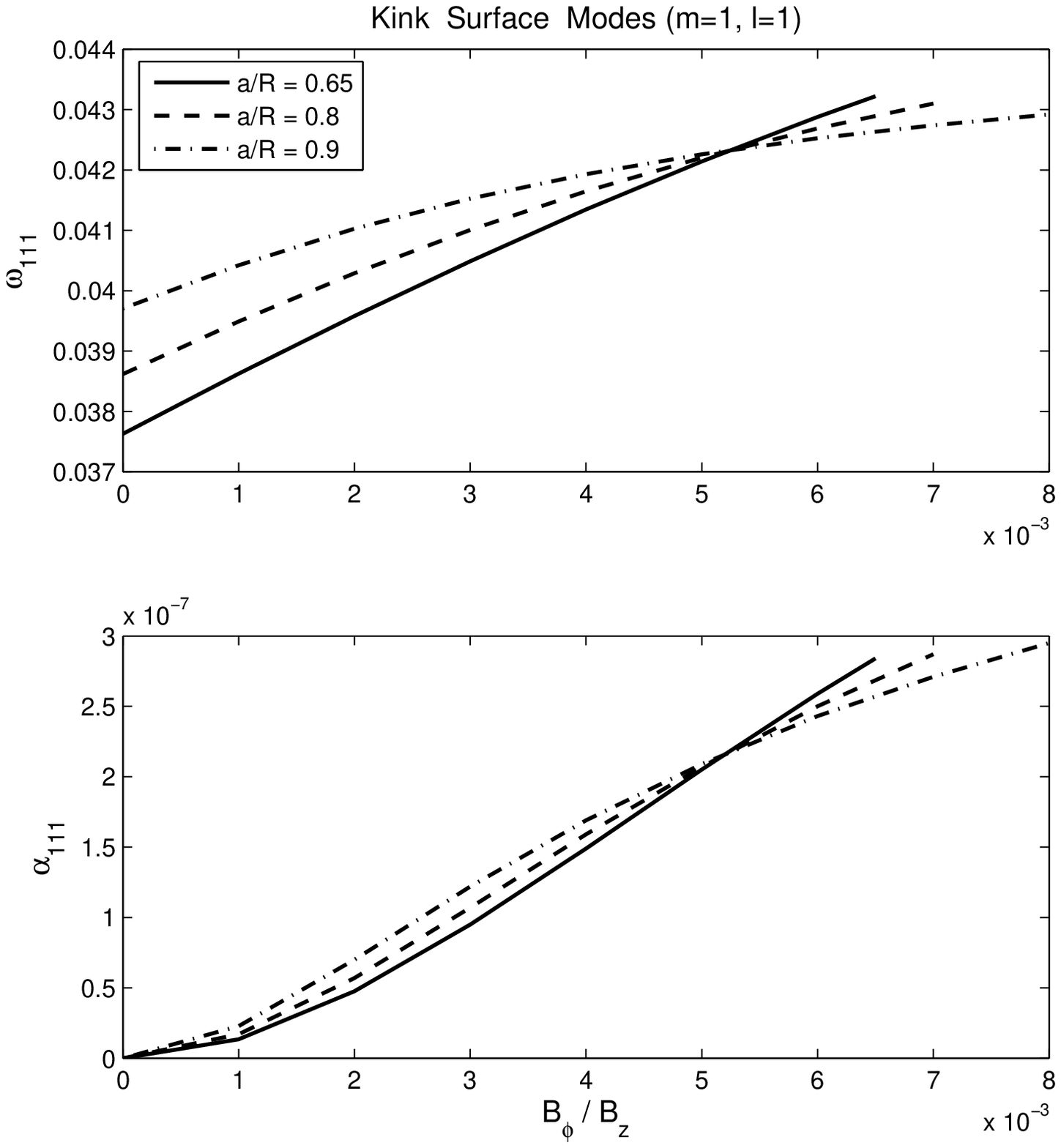}
      \vspace{10.1cm}
      \caption[]{Frequency of the fundamental kink ($m=1$) surface mode and its damping
rate versus the twist parameter, $B_{\phi}/B_z$, for different
relative core width $a/R=$ 0.65 (solid), 0.8 (dashed) and 0.9
(dash-dotted). The loop parameters are: $L=109\times 10^3$ km,
$R/L=0.01$, $\rho_{e}/\rho_{i}=0.1$, $\rho_{0}/\rho_{i}=0.5$,
$\rho_{i}=2\times 10^{-14}$ gr cm$^{-3}$, $B_{e}/B_{i}=1$,
$B_{0}/B_{i}=1$, $B_{i}=100$ G, $S=10^4$ and ${\mathcal{R}}=560$.
Both frequencies and damping rates are in units of the interior
Alfv\'{e}n frequency, $\omega_{\rm A_i}=1.835{\rm~rad~s^{-1}}$.}
         \label{m1l1-surface-wa}
   \end{figure}

\begin{figure}
\center \includegraphics{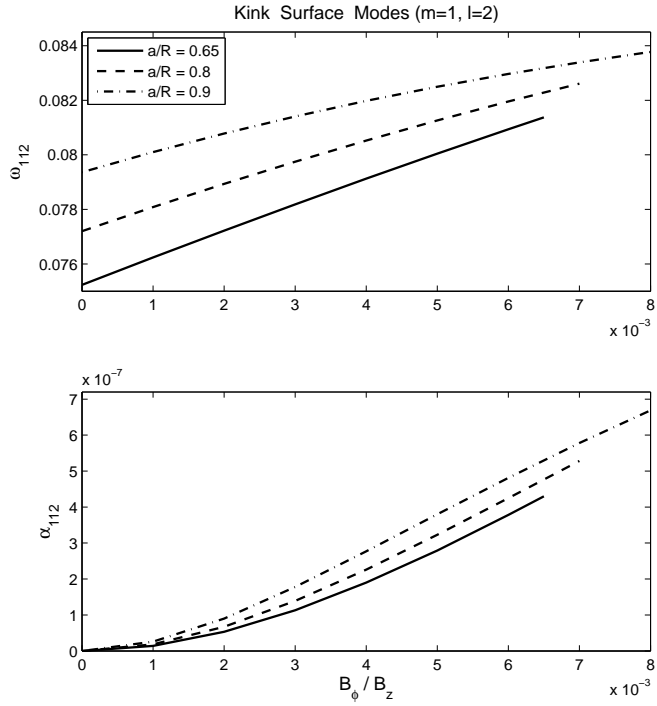}
      \vspace{10.1cm}
      \caption[]{Same as Fig. \ref{m1l1-surface-wa}, for the first-overtone kink ($m=1$) surface modes.}
         \label{m1l2-surface-wa}
   \end{figure}

\begin{figure}
\center \includegraphics{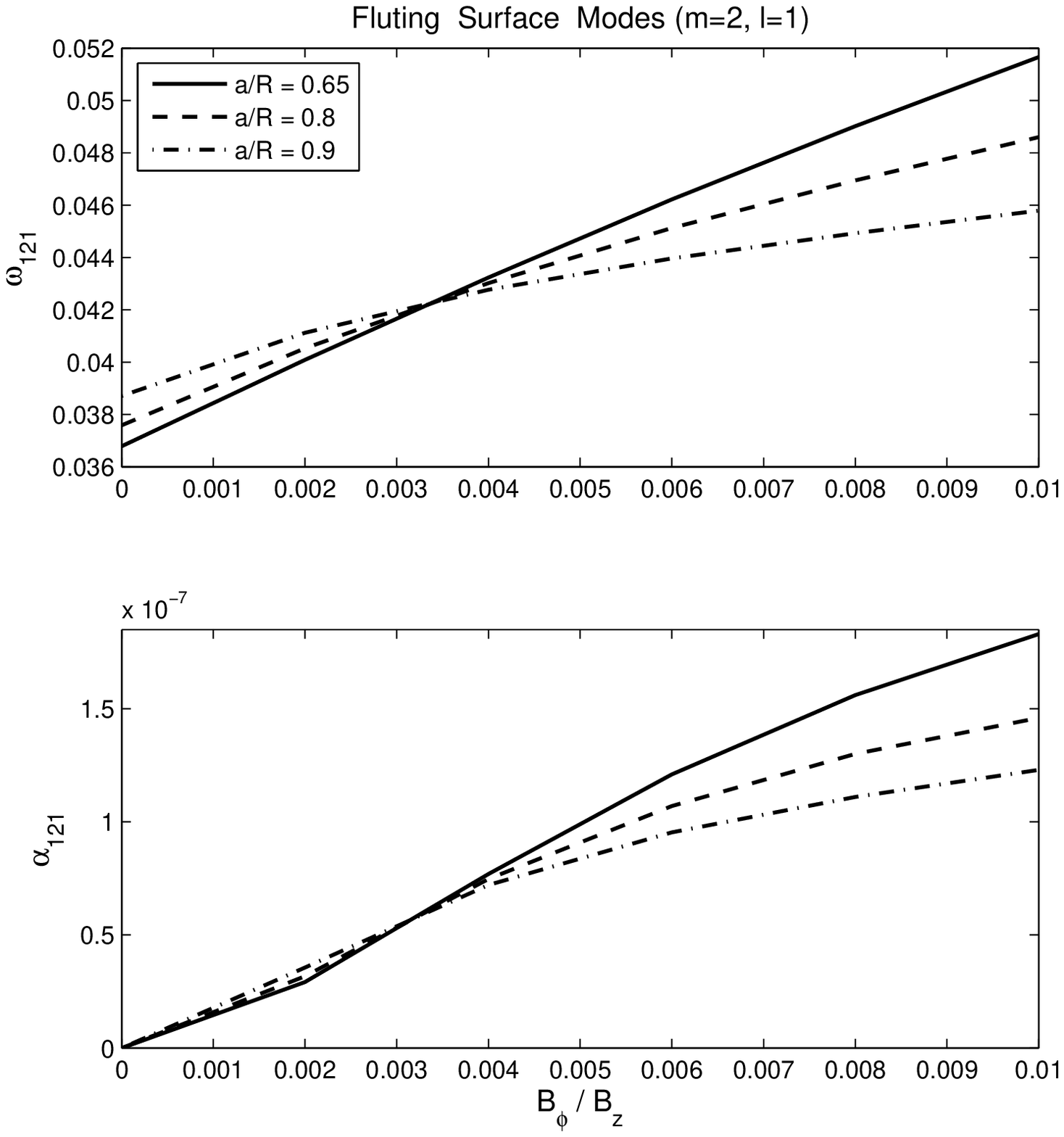}
      \vspace{10.1cm}
      \caption[]{Same as Fig. \ref{m1l1-surface-wa}, for the fundamental fluting ($m=2$) surface modes.}
         \label{m2l1-surface-wa}
   \end{figure}

\begin{figure}
\center \includegraphics{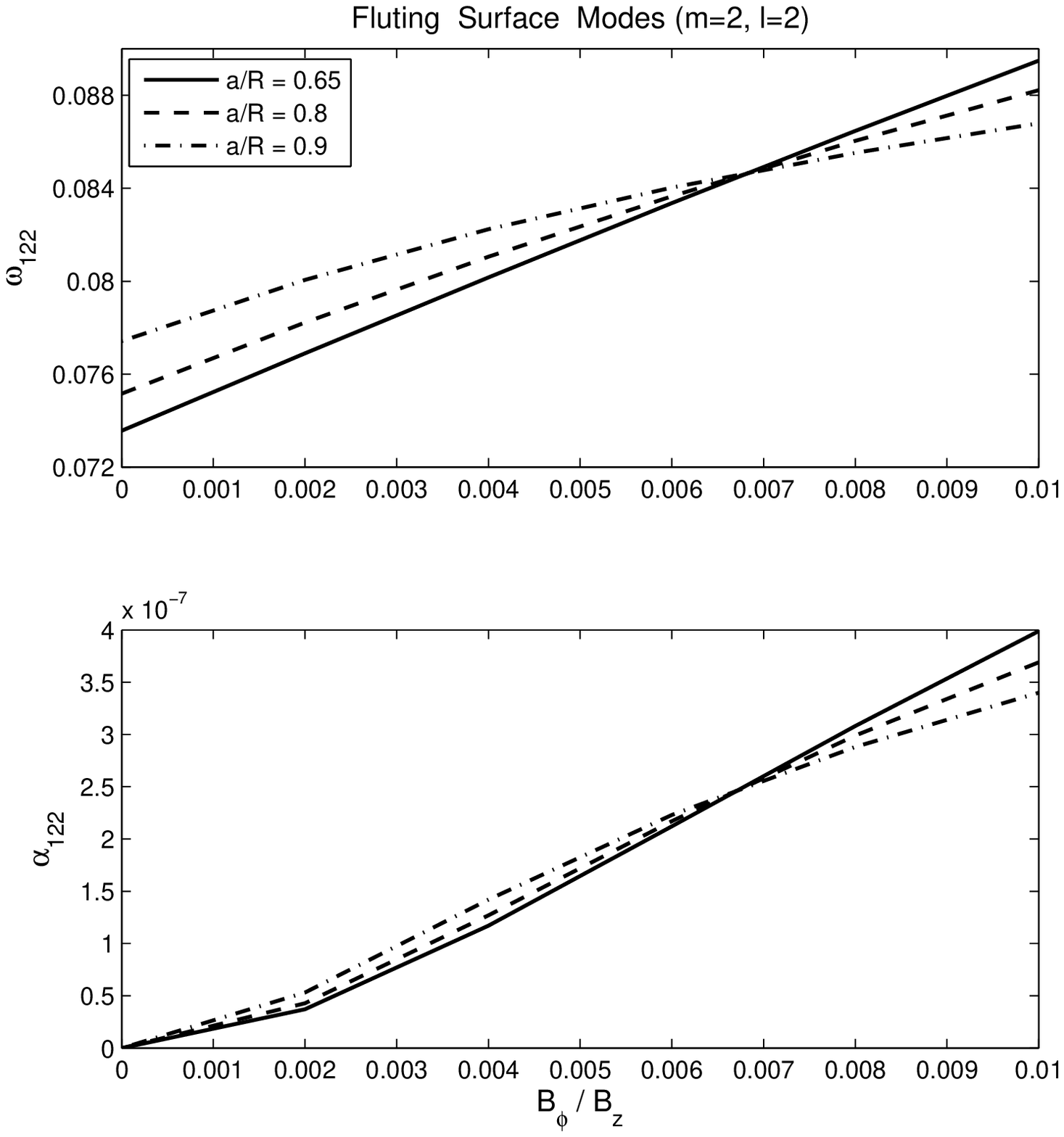}
      \vspace{10.1cm}
      \caption[]{Same as Fig. \ref{m1l1-surface-wa}, for the first-overtone fluting ($m=2$) surface modes.}
         \label{m2l2-surface-wa}
   \end{figure}

\begin{figure}
\center \includegraphics{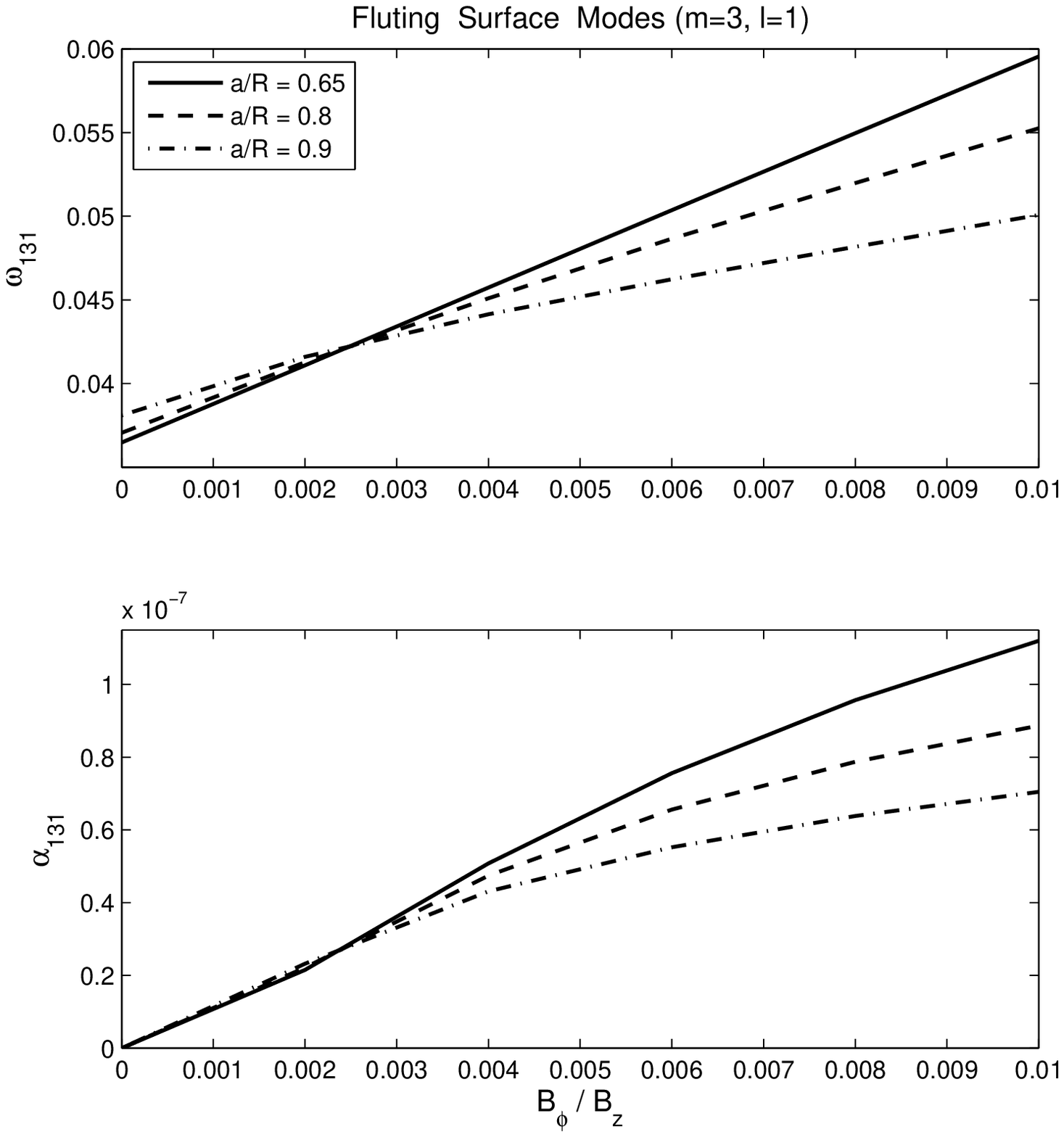}
      \vspace{10.1cm}
      \caption[]{Same as Fig. \ref{m1l1-surface-wa}, for the fundamental fluting ($m=3$) surface modes.}
         \label{m3l1-surface-wa}
   \end{figure}

\begin{figure}
\center \includegraphics{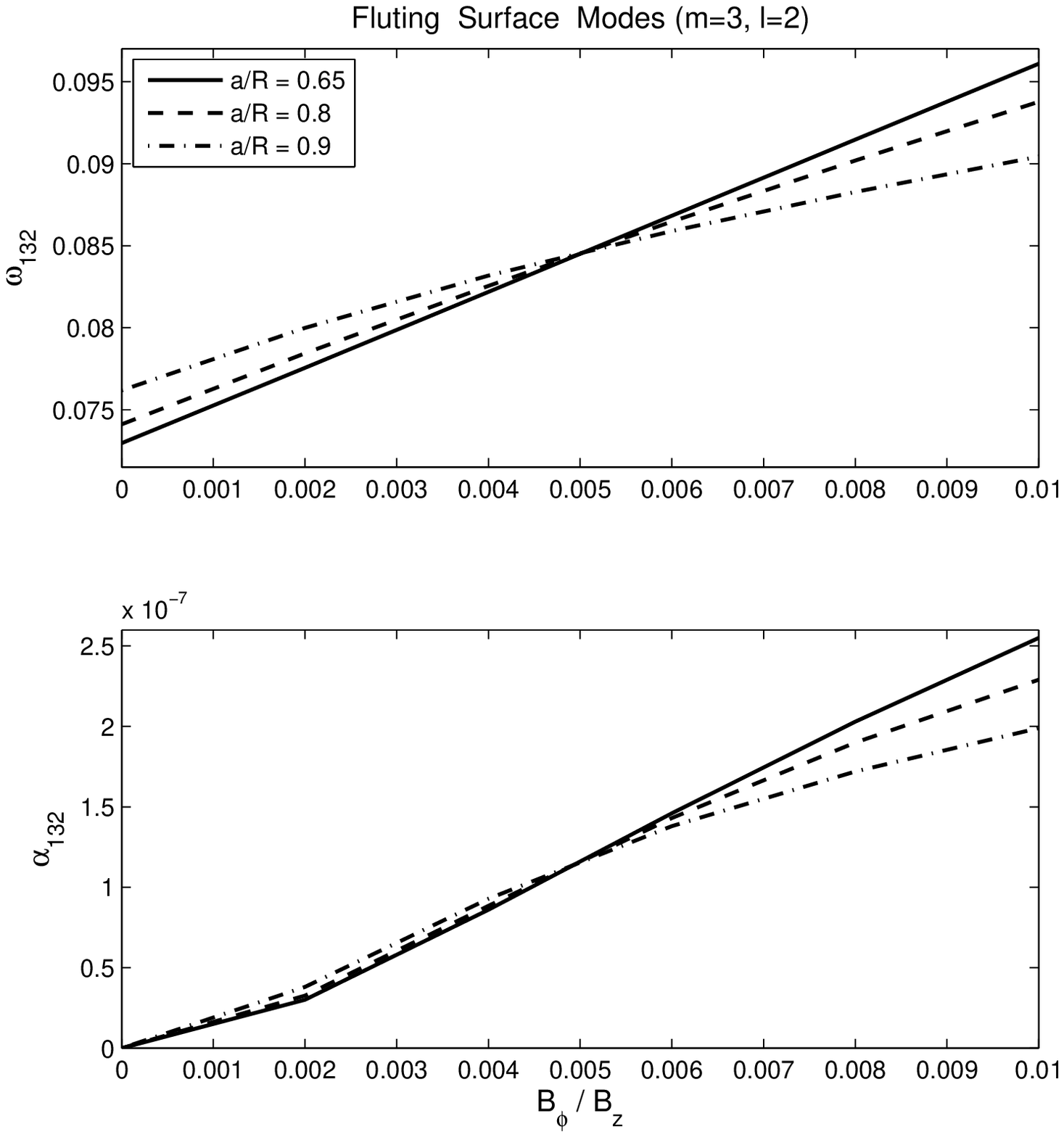}
      \vspace{10.1cm}
      \caption[]{Same as Fig. \ref{m1l1-surface-wa}, for the first-overtone fluting ($m=3$) surface modes.}
         \label{m3l2-surface-wa}
   \end{figure}

\begin{figure}
\center \includegraphics{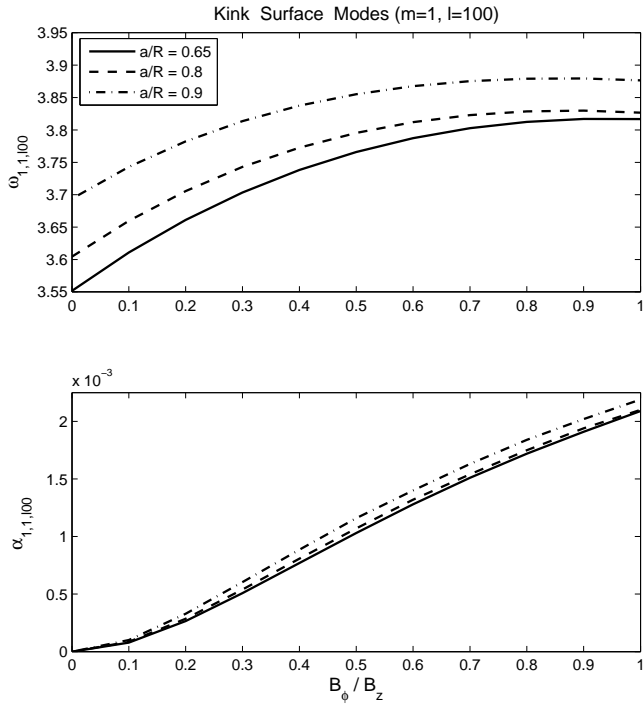}
      \vspace{10.1cm}
      \caption[]{Same as Fig. \ref{m1l1-surface-wa}, for the kink ($m=1$) surface modes with $l=100$.}
         \label{m1l100-surface-wa}
   \end{figure}

The effects of twisted magnetic field on both the frequencies
$\omega$ and damping rates $\alpha$ are calculated by numerical
solution of the dispersion relation, i.e. Eqs. (28a)-(28b) given by
Erd\'{e}lyi \& Carter (2006), and Eq. (\ref{alpha}), respectively.
The results are displayed in Figs. \ref{m1l1-surface-wa} to
\ref{m1l100-surface-wa}. Figures \ref{m1l1-surface-wa} to
\ref{m3l2-surface-wa} show the frequencies and damping rates of the
fundamental and first-overtone $l=1,2$ kink $(m = 1)$ and fluting
$(m=2,3)$ surface modes versus the twist parameter,
$B_{\phi}/B_z:=\frac{A_0R}{B_0}$, and for different relative core
width $a/R=(0.65, 0.8, 0.9)$. Figures \ref{m1l1-surface-wa} to
\ref{m3l2-surface-wa} reveal that: i) For a given $a/R$, both
frequencies and their corresponding damping rates increase when the
twist parameter increases. The result of $\omega$ is in good
agreement with the that obtained by Carter $\&$ Erd\'{e}lyi (2008).
Note that there are two surface modes labelled by $(n=1,2)$ which
are in accordance with Carter $\&$ Erd\'{e}lyi (2008). Here we only
show the first one $(n=1)$ in the figures, because the second one
$(n=2)$ does not show itself in all selected twists. ii) For a given
$m$ and $a/R$, when the longitudinal mode number, $l$, increases,
both the frequencies and damping rates increase. iii) For a given
$l$, $a/R$ and $B_{\phi}/B_z$, when the azimuthal mode number, $m$,
increases, the frequencies and damping rates increase and decrease,
respectively.

Figure \ref{m1l100-surface-wa} presents the frequencies and damping
rates of the kink $(m = 1)$ surface modes with $l=100$ versus the
twist parameter. Figure \ref{m1l100-surface-wa} shows that for
$l=100$, the damping becomes stronger and the ratio $\omega/\alpha$
decreases two order of magnitude compared with $l=1,2$. See again
Figs. \ref{m1l1-surface-wa} to \ref{m1l2-surface-wa}.

Here in our calculations, the sausage modes ($m=0$) are absent.
Because following Edwin \& Roberts (1983) and Karami, Nasiri $\&$
Sobouti (2002), the sausage modes have a lower longitudinal cutoff
and they are only expected in fat and dense loops. For instance,
according to Aschwanden (2005) for typical active region loops which
have a density contrast in the order of $\rho_{\rm e}/\rho_{\rm
i}\approx0.1-0.5$, would be required to have width-to-length ratios
of $L/(2R)\approx 1-2$.

\begin{figure}
\center \includegraphics{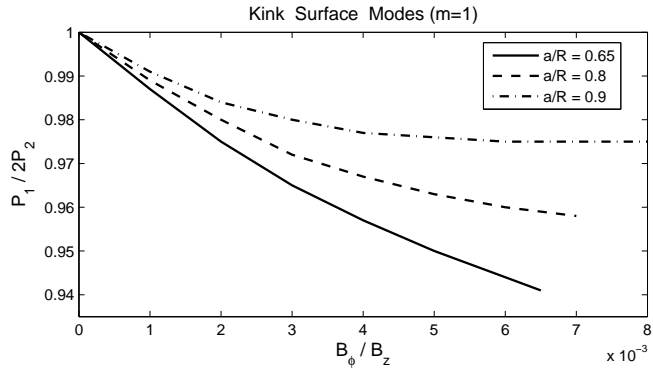}
      \vspace{8.0cm}
      \caption[]{The period ratio $P_1/2P_2$ of the fundamental and its first-overtone kink ($m=1$) surface modes versus the twist parameter
  for different relative core width $a/R=$ 0.65 (solid), 0.8 (dashed) and
0.9 (dash-dotted). Auxiliary parameters as in Fig.
\ref{m1l1-surface-wa}.}
         \label{m1-surface-p1p2}
   \end{figure}

\begin{figure}
\center \includegraphics{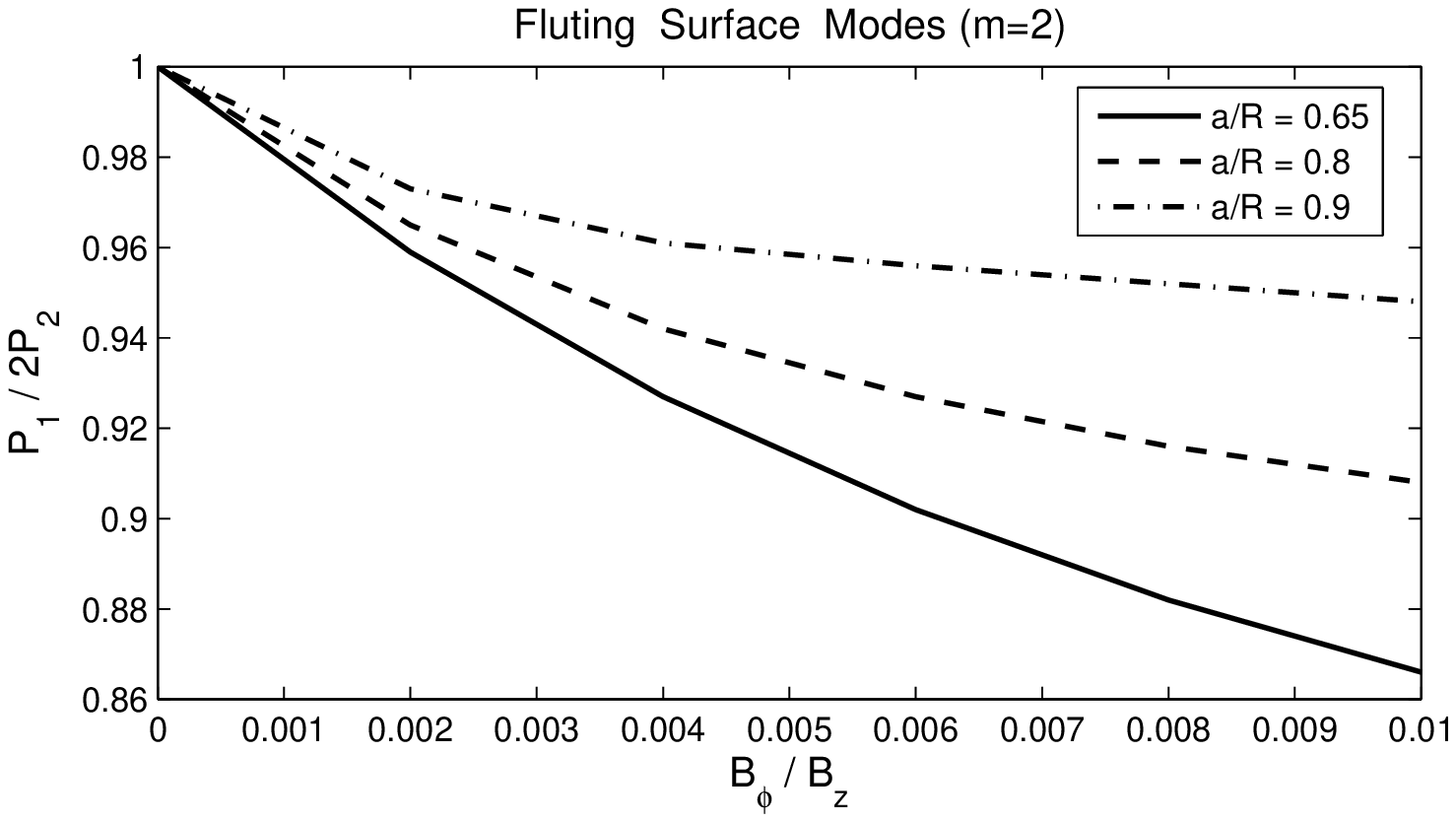}
      \vspace{8.0cm}
      \caption[]{Same as Fig. \ref{m1-surface-p1p2}, for fluting ($m=2$) surface modes.}
         \label{m2-surface-p1p2}
   \end{figure}

\begin{figure}
\center \includegraphics{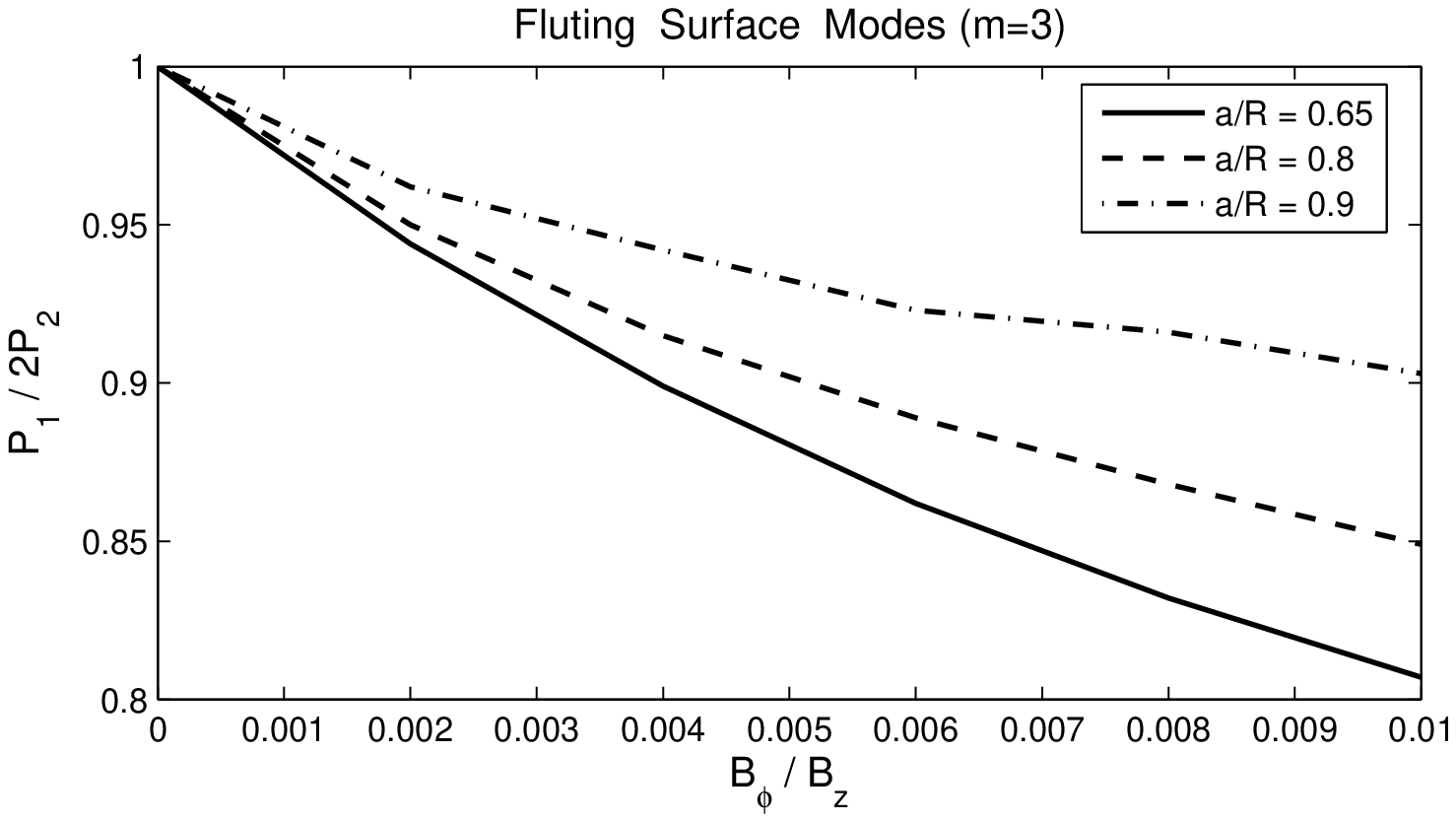}
      \vspace{8.0cm}
      \caption[]{Same as Fig. \ref{m1-surface-p1p2}, for fluting ($m=3$) surface modes.}
         \label{m3-surface-p1p2}
   \end{figure}

The period ratio $P_1/2P_2$ of the fundamental and first-overtone,
$l=1,2$ modes of both the kink ($m=1$), and fluting ($m=2,3$)
surface waves versus the twist parameter plotted in Figs.
\ref{m1-surface-p1p2} to \ref{m3-surface-p1p2}, respectively.
Figures \ref{m1-surface-p1p2} to \ref{m3-surface-p1p2} show that: i)
For a given relative core width, the period ratio $P_1/2P_2$
decreases when the twist parameter increases. For $a/R$=0.65, for
instance, $P_1/2P_2$ decrease from 1 (for untwisted loop) and
approaches below 0.95, 0.88 and 0.82 for $m=(1,2,3)$, respectively,
with increasing the twist parameter. ii) For a given twist
parameter, the period ratio $P_1/2P_2$ increases when the relative
core width increases. Figure \ref{m1-surface-p1p2} clears that for
kink modes $(m=1)$ with $B_{\phi}/B_z$=0.0065 and $a/R=0.65$, the
ratio $P_1/2P_2$ is 0.941. This is in good agreement with the period
ratio observed by Verwichte et al. (2004), 0.91$\pm$0.04 deduced
from the observations of TRACE. See also McEwan, D\'{i}az $\&$
Roberts (2008).

\begin{figure}
\center \includegraphics{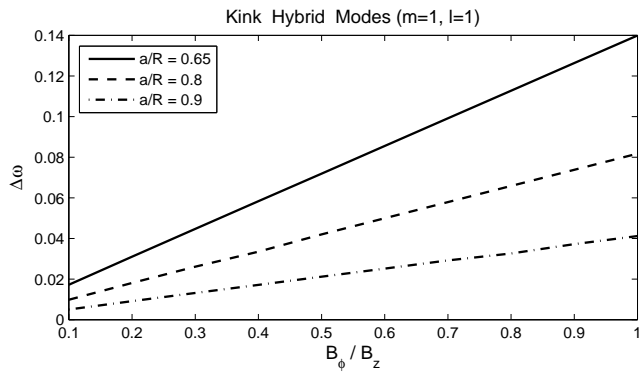}
      \vspace{8.0cm}
      \caption[]{Frequency band width of the fundamental kink ($m=1$) hybrid modes versus the twist parameter
       for different relative core width $a/R=$ 0.65 (solid), 0.8 (dashed) and
0.9 (dash-dotted). Auxiliary parameters as in Fig.
\ref{m1l1-surface-wa}.}
         \label{m1-hybrid-dw}
   \end{figure}

Figure \ref{m1-hybrid-dw} displays the frequency band width,
$\Delta\omega$, including infinite set of the fundamental kink
$(m=1)$ hybrid modes versus the twist parameter and for different
relative core width. Figure \ref{m1-hybrid-dw} presents that: i) For
a given twist parameter, $\Delta\omega$ increases when the relative
core width decreases. ii) For a given relative core width,
$\Delta\omega$ increases when the twist parameter increases. This is
in good agreement with the result obtained by Carter $\&$
Erd\'{e}lyi (2008).

\section{Conclusions}
Oscillations and damping of standing fast MHD surface and hybrid
waves in coronal loops in presence of twisted magnetic field is
studied. To do this, a typical coronal loop is considered as a
straight pressureless cylindrical incompressible flux tube with
magnetic twist just in the annulus and straight magnetic field in
the internal and external regions.  The linearized MHD equations,
when the dissipation is absent, are reduced to a Bessel's equation
for the perturbed magnetic pressure. The dispersion relation is
obtained and solved numerically for obtaining the frequencies of
both the kink and fluting modes. The damping rates of oscillations
due to the resistive and viscous dissipation in presence of the
magnetic twist is obtained and solved numerically. Our numerical
results show that:

i) For a given relative core width, frequencies and damping rates of
both the kink $(m = 1)$ and fluting $(m=2,3)$ surface waves increase
when the twist parameter increases.

ii) The period ratio $P_1/2P_2$, for both the kink ($m=1$) and
fluting ($m=2,3$) surface modes are lower than 1 (for untwisted
loop) in presence of the twisted magnetic field. The result of
$P_1/2P_2$ for kink modes is in accordance with the TRACE
observations.

iii) Frequency band width of the fundamental kink ($m=1$) hybrid
modes increase when the twist parameter increases.

\section*{Acknowledgments}
This work was supported by the Research Institute for Astronomy
$\&$ Astrophysics of Maragha (RIAAM), Maragha, Iran.


\clearpage

\label{lastpage}

\end{document}